\documentclass[aps,prl,preprint,groupedaddress]{revtex4-1}

\usepackage{graphicx}
\graphicspath{ {./Figures/} }
\usepackage{epstopdf}
\usepackage{float}
\usepackage{color}

\begin{document}


\title{Correlation between Superconductivity, Band Filling and Electron Confinement at the LaAlO$_{3}$-SrTiO$_{3}$ Interface}


\author{A.E.M. Smink}
\email{a.e.m.smink@utwente.nl}
\author{M.P. Stehno}
\author{J.C. de Boer}
\author{A. Brinkman}
\author{W.G. van der Wiel}
\author{H. Hilgenkamp}
\affiliation{MESA+ Institute for Nanotechnology, University of Twente, P.O. Box 217, 7500 AE Enschede, The Netherlands}


\date{\today}

\begin{abstract}
By combined top- and backgating, we explore the correlation of superconductivity with band filling and electron confinement at the LaAlO$_3$-SrTiO$_3$ interface. We find that the top- and backgate voltages have distinctly different effects on the superconducting critical temperature, implying that the confining potential well has a profound effect on superconductivity. We investigate the origin of this behavior by comparing the gate-dependence of $T_c$ to the corresponding evolution of the band filling with gate voltage. For several backgate voltages, we observe maximum $T_c$ to consistently coincide with a kink in tuning the band filling for high topgate voltage. Self-consistent Schr\"odinger-Poisson calculations relate this kink to a Lifshitz transition of the second $d_{xy}$ subband. These results establish a major role for confinement-induced subbands in the phase diagram of SrTiO$_3$ surface states, and establish gating as a means to control the relative energy of these states.
\end{abstract}


\maketitle


Electron-doped strontium titanate (SrTiO$_3$) is the first oxide and the first semiconductor reported to become superconducting \cite{schooley_superconductivity_1964}, stimulating many research efforts to understand and utilize this superconductivity. Bulk SrTiO$_3$ can be doped either through reduction by formation of oxygen vacancies \cite{schooley_superconductivity_1964}, or by cation substitution \cite{suzuki_superconductivity_1996, koonce_superconducting_1967}. With a maximum superconducting critical temperature $T_c$ around 400 mK, bulk SrTiO$_3$ superconductivity persists down to carrier densities as low as $10^{17} $ cm$^{-3}$ \cite{schooley_dependence_1965, lin_critical_2014}. Besides by bulk doping, superconductivity has also been achieved in the quasi-two-dimensional electron system (q-2DES) formed at the surface of stoichiometric SrTiO$_3$, by either ionic-liquid gating \cite{ueno_electric-field-induced_2008} or by deposition of a selected overlayer such as LaAlO$_3$ \cite{reyren_superconducting_2007}.

In these surface states, superconductivity is two-dimensional with an in-plane superconducting coherence length of $\sim$50 nm and a thickness of $\sim$10 nm \cite{reyren_anisotropy_2009}. The superfluid density is on the order of $10^{11}$ to $10^{12}$ cm$^{-2}$ \cite{bert_direct_2011}, enabling electrostatic control of the superconducting state, a major topic in correlated electron physics \cite{ahn_electric_2003}. This was demonstrated almost simultaneously on bare SrTiO$_3$ surfaces by ionic-liquid gating \cite{ueno_electric-field-induced_2008}, and at the interface between LaAlO$_3$ and SrTiO$_3$ by backgating through the insulating SrTiO$_3$ substrate \cite{caviglia_electric_2008}. Using the LaAlO$_3$ layer as gate dielectric (topgating), the latter system was used for MOSFET-like devices to locally switch superconductivity \cite{eerkes_modulation_2013} and to create devices with novel functionality \cite{cheng_electron_2015, goswami_quantum_2016}.

In many unconventional superconductors, $T_c$ has a dome-like dependence on an externally controlled parameter, for example hydrostatic pressure \cite{takabayashi_disorder-free_2009, slooten_enhancement_2009} and doping by chemical \cite{lee_doping_2006, schaak_superconductivity_2003} or electrostatic \cite{mannhart_high-Tc_1996, ye_superconducting_2012} means. Both in the bulk and in surface states of SrTiO$_3$, a comparable dependence of $T_c$ on either chemical or electrostatic doping was revealed \cite{schooley_dependence_1965, lin_critical_2014, caviglia_electric_2008, ueno_electric-field-induced_2008}, showing similarities to other unconventional superconductors. At SrTiO$_3$-based interfaces, the low superfluid density should enable exploration of this entire phase diagram using electrostatic gating.

In such gating experiments, the maximum $T_c$ was reported to occur at different values for the carrier density $n_{2D}$ \cite{joshua_universal_2012, biscaras_two-dimensional_2012, hurand_field-effect_2015}, suggesting that $n_{2D}$ is not the sole factor determining the phase diagram. This led to proposals to base the phase diagram on the sheet conductivity \cite{biscaras_multiple_2013, fete_large_2014, gariglio_interface_2015, prawiroatmodjo_evidence_2016}, which also does not provide a universal result. Almost all these experiments were done in a backgate geometry, whereas topgating has a different effect on carrier mobility \cite{bell_dominant_2009, hosoda_transistor_2013, chen_dual-gate_2016} and on the band structure \cite{smink_gate-tunable_2017}. This difference is due to the opposite direction of the applied electric field, resulting in a disparate effect on the shape of the confining potential well. A combination of both gating geometries would allow to control separately both the carrier density and the shape of the potential well, revealing their individual effects on superconductivity at SrTiO$_3$-based interfaces.

In this work, we explore the effect of simultaneous top- and backgating on superconductivity and on the band filling at the (001) LaAlO$_3$-SrTiO$_3$ interface. We reveal a striking asymmetry in the top- and backgate dependence of $T_c$, indicating that the shape of the confining potential well strongly affects superconductivity at the surface of SrTiO$_3$. We investigate this effect further by measuring the corresponding effect of both gate voltages on the band filling, in subsequent magnetotransport experiments above $T_c$. In these measurements, we demonstrate tuning the carrier density of the $d_{xz,yz}$ Lifshitz transition, and tuning of the topgate-dependent superconducting dome by a backgate voltage. At the topgate voltage where $T_c$ is maximized, we observe a kink in the gate-dependence of the $d_{xy}$ carrier density. By Schr\"odinger-Poisson calculations, we attribute this kink to depleting the second $d_{xy}$ subband with increasing carrier density.

The fabrication of the topgated Hall bar devices is described in Ref. \cite{smink_gate-tunable_2017}. Here, we present the results for a 50 $\mu$m wide Hall bar; a second device showed similar behavior. All measurements were performed in a dilution refrigerator with 10 mK base temperature, using a lock-in amplifier with an excitation current of 1 nA, far below the critical current for superconductivity in our samples ($\sim$500 nA). The topgate leakage current was kept below 100 pA during the measurements, the backgate leakage current was always below the measurement limit of $\sim$1 pA.

The gate voltages were applied with respect to the grounded current drain, and the silver paste gluing the sample to a copper plate served as the backgate electrode. To ensure reproducible gate sweeps \cite{biscaras_limit_2014}, the topgate (backgate) voltage was swept to +1.5 V (0 V), to -0.7 V (-20 V), and back to 0 V prior to measurement, at $T$ = 500 mK. During measurement, the topgate voltage was always swept from positive to negative values. Between measurements, the zero-gate-voltage data were measured several times, which always overlapped with the curve measured at the start of the experiment. All $R(T)$ curves were taken first, after which the magnetotransport was measured above $T_c$, at $T$ = 500 mK.

Like recently reported for modulation-doped SrTiO$_3$ interfaces \cite{trier_quantization_2016}, we find that the SrCuO$_2$ capping enhances the effect of a backgate voltage compared to samples without this capping. Both modulation doping and SrCuO$_2$ capping suppress the formation of scattering centers at the interface, which increases the mobility \cite{huijben_defect_2013, chen_extreme_2015}. In samples with a higher density of scattering centers, these impurities can screen the electric field of the backgate, thus suppressing its gate effect. In our samples, the enhanced gate effect has an important implication. The gate-voltage range is limited because the contacts become insulating already below a backgate voltage of -20 V. Compared to SrCuO$_2$-capped interfaces without a topgate, we also find that depositing the Au topgate electrode reduces the mobility and raises the carrier density to values reported for uncapped LaAlO$_3$-SrTiO$_3$ interfaces \cite{bell_dominant_2009, hosoda_transistor_2013}.

\begin{figure}
\includegraphics[width=1\columnwidth]{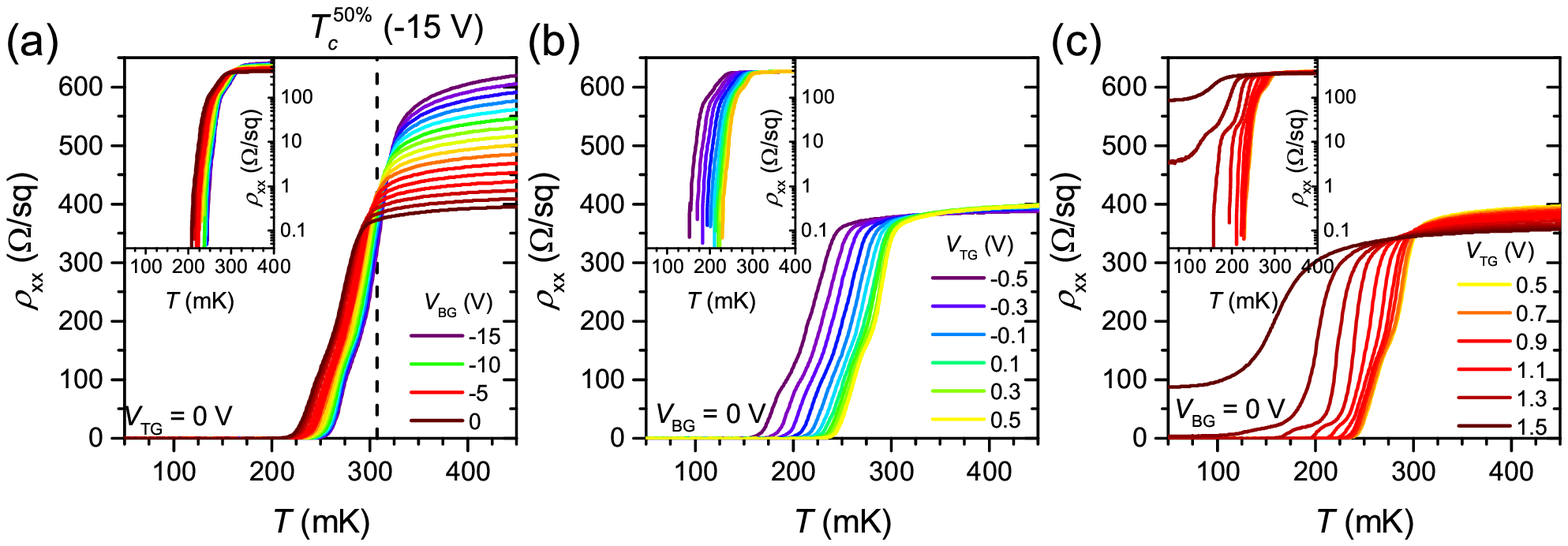}
\caption{Tuning the superconducting transition with individual top- and backgate voltages. Resistivity versus temperature as function of (a) backgate voltage, (b) topgate voltage below the point where $T_c$ is maximized, (c) topgate voltage above this point. Insets: semilogarithmic plots of the same data, showing more clearly the multistep transition. In (a), the extraction of $T_c$ for a backgate voltage of -15 V is illustrated by the dashed line.}
\label{RTBGTG}
\end{figure}

Figure \ref{RTBGTG} shows the effect of an individual topgate ($V_{TG}$) or backgate ($V_{BG}$) voltage on the superconducting transition upon cooldown, with the other gate voltage set to 0 V. Figures \ref{RTBGTG}(a) and (b) show that the two gate voltages have an opposite effect on the transition temperature. From the ungated situation ($V_{BG}=V_{TG}=0$ V), the transition shifts to higher temperature with increasing topgate voltage, or with decreasing backgate voltage. A positive voltage on either gate increases the carrier density at the interface, so the total 2D carrier density cannot be the sole factor determining superconductivity at the LaAlO$_3$-SrTiO$_3$ interface. Instead, the difference between top- and backgating suggests that details of the electrostatics play an important role.

Above $V_{TG} =$ +0.5 V, the transition temperature starts to decrease and the shape of the $R(T)$ curve changes considerably. It shows multiple steps towards the zero resistance state, and a partial transition for the highest topgate voltages. This behavior indicates multiple superconducting transitions, suggesting a percolative superconducting transition
resulting from inhomogeneity \cite{caprara_effective_2011, poccia_percolative_2014, eley_approaching_2011}. For SrTiO$_3$-based q-2DESs, inhomogeneity due to electronic phase separation is predicted to be an intrinsic property \cite{ariando_electronic_2011, scopigno_phase_2016}, depending on an applied gate voltage \cite{scopigno_phase_2016, singh_competition_2018}. Another property that can cause inhomogeneity at the surface of SrTiO$_3$ is tetragonal domain formation with gate voltage, which drives local variations in current density and critical temperature \cite{honig_local_2013, kalisky_locally_2013, noad_variation_2016, ma_local_2016}.

\begin{figure}
\includegraphics[width=0.5\columnwidth]{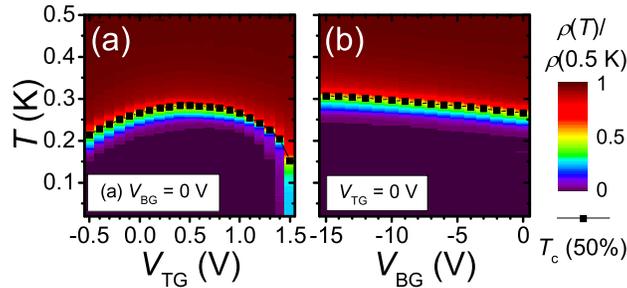}
\caption{Tuning of $T_c$ by combined top- and backgating. (a)-(b) Color plot of normalized resistance versus temperature as function of individual (a) topgate voltage and (b) backgate voltage. The critical temperature, extracted as described in the text, is indicated by the black line.
}
\label{TCBGTG}
\end{figure}

We observe the resistive-transition steps to be close together in temperature for all gate voltages. In the remainder of this paper, we therefore omit the details of the transition and use the global transition temperature $T_c$ to describe the effect of the gate voltages on superconductivity. We define $T_c$ through the relation $R(T_c) = 0.5 \times R$(500 mK). Figure \ref{TCBGTG}(a) shows a dome-like dependence of $T_c$ on the topgate voltage, in line with previous experiments \cite{caviglia_electric_2008, joshua_universal_2012, richter_interface_2013, hurand_field-effect_2015}. The backgate data in Fig. \ref{TCBGTG}(b) do not show a full dome-like dependence of $T_c$, but a qualitative comparison to previous results \cite{caviglia_electric_2008, joshua_universal_2012} suggests that $T_c$ would be maximized just below the minimum gate voltage of this measurement.

To better understand the gate tuning of $T_c$, we now investigate the effect of simultaneous top- and backgating on the band filling. The carrier density and mobility were extracted from magnetotransport data, see the Supplemental Material for details. Like in Refs. \cite{joshua_universal_2012, biscaras_two-dimensional_2012, chen_dual-gate_2016, smink_gate-tunable_2017}, fitting the magnetotransport data required using two carrier types with distinct mobility. At the lowest gate voltages, only one carrier type can be distinguished. Since the $d_{xy}$ band lies lower in energy than the $d_{xz,yz}$ bands \cite{salluzzo_orbital_2009, santander-syro_two-dimensional_2011}, these carriers are most likely of the $d_{xy}$ type; the other carriers reside in the $d_{xz,yz}$ bands.

\begin{figure}
\includegraphics[width=0.5\columnwidth]{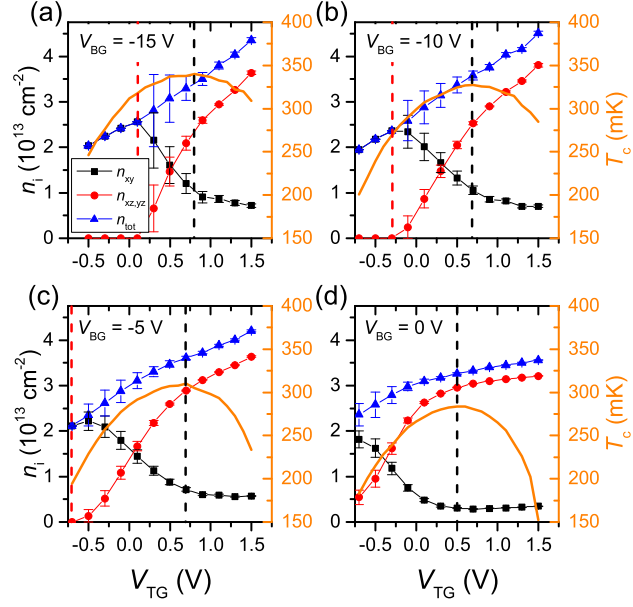}
\caption{Comparison of the evolution of $T_c$ and band filling with topgate voltage, for varying backgate voltage. The applied backgate voltage is (a) -15 V, (b) -10 V, (c) -5 V, and (d) 0 V. Lines connecting data points are guides to the eye. For ease of comparison, all axes have the same limits. The dashed, vertical lines indicate the characteristic topgate voltages: the black one marks the topgate voltage where $T_c$ is maximized, the red line indicates the $d_{xz,yz}$ Lifshitz transition.}
\label{comp}
\end{figure}

Figure \ref{comp} displays the topgate-voltage dependence of the carrier density per band (band filling of $d_{xy}$ and $d_{xz,yz}$), for four different backgate voltages. In all panels, we observe $d_{xz,yz}$ carriers to start contributing to transport upon increasing the topgate voltage. This emergence of $d_{xz,yz}$ carriers marks the appearance of additional electron pockets in the Fermi surface. Such a change in the topology of the Fermi surface defines a Lifshitz transition \cite{yamaji_quantum_2006}. Gate tuning through this $d_{xz,yz}$ Lifshitz transition has been firmly established: it has been reported for several SrTiO$_3$-based interfaces, using either back- \cite{joshua_universal_2012, biscaras_two-dimensional_2012} or topgating \cite{smink_gate-tunable_2017, niu_giant_2017}. However, since these articles report different values for the corresponding carrier density, referred to as the $d_{xz,yz}$ Lifshitz density of $n_L$ below, it may depend on other factors than the band structure alone. Note from Fig. \ref{comp}, that superconductivity persists far below $n_L$, where only the $d_{xy}$ subbands are populated.

In line with Refs. \cite{smink_gate-tunable_2017, niu_giant_2017}, Figure \ref{comp} shows that $n_{xy}$ decreases when the $d_{xz,yz}$ bands start to be populated. In a simple model first proposed by Maniv \textit{et al.} \cite{maniv_strong_2015}, this behavior is attributed to electron-electron interactions. In the model, these interactions are proposed as a Hubbard-type repulsion between electrons in different orbitals in the same unit cell. Therefore, the strength of these interactions is modeled as a phenomenological Coulomb screening parameter, $U$. The interactions push $d_{xy}$ subbands, with a lower density of states (DOS), upwards in energy when a band with large DOS ($d_{xz,yz}$) crosses the Fermi level. This results in a strong decrease of $n_{xy}$ upon increasing the total carrier density; these carriers are redistributed into the $d_{xz,yz}$ bands.

Based on Fig. \ref{comp}, we can directly compare the effect of both gate voltages on the band filling to the corresponding evolution of $T_c$. For the latter, we observe that the backgate voltage affects the topgate dependence of $T_c$ in shape, height, and peak position. In the following, we focus on the gate effect on the peak position, because it marks the conditions for optimal superconductivity.

A closer look at the topgate dependence of the band filling around this peak position reveals a surprising feature: there is a kink in tuning the carrier density per band with topgate voltage. This kink is most pronounced for $V_{BG} = -15$ V. Therefore, we will focus first on the results for this backgate voltage, and consider the effect of changing the backgate voltage afterwards. To gain insight into the origin of this kink, we performed self-consistent Schr\"odinger-Poisson calculations, using a slight adaptation of the code used in Ref. \cite{smink_gate-tunable_2017}. The two adaptations made are (i) changing the thickness of the bound background charge layer to 50 nm, and (ii) adding the effect of a backgate voltage as described in the Supplemental Material. 

Figure \ref{SP_comp}(a) shows the result of these calculations for the band filling as function of the total carrier density, for a backgate voltage of -15 V, a background charge density $n_b = 6.1 \times 10^{13}$ cm$^{-2}$, and Coulomb screening parameter $U = 1.8$ eV. This background charge density is in good agreement with thermodynamic approaches to defect chemistry \cite{gunkel_space_2016, gunkel_mobility_2017} and with previous Schr\"odinger-Poisson calculations \cite{gariglio_electron_2015}. The Coulomb screening parameter also corresponds well with previous reports \cite{maniv_strong_2015, breitschaft_two-dimensional_2010}. We find a remarkably good fit between the experimental data and the calculations, reproducing both the $d_{xz,yz}$ Lifshitz transition and the kink in the filling coinciding with maximum $T_c$. For the other backgate voltages, using the same parameters results in reasonable fits, which are discussed in the Supplemental Material. Based on the quality of these fits, we take the results of these calculations as the basis for further discussion of our experimental data.

Both in the experimental results and in the calculations, for $V_{BG} = -15$ V, the kink feature occurs at a total carrier density of $\sim 3.5 \times 10^{13}$ cm$^{-2}$. Using the Schr\"odinger-Poisson calculations, we can investigate the band structure for a total carrier density just below and just above this point. Panels (b)-(d) of Fig. \ref{SP_comp} show the calculated potential well, its bound states and the band dispersions along $k_x$, for $n_{tot} = 3.39 \times 10^{13}$ cm$^{-2}$ (b)-(c) and $3.59 \times 10^{13}$ cm$^{-2}$ (d). A comparison of panel (c) to panel (d) shows that the second-order subband of the $d_{xy}$ type is pushed above the Fermi level at this point. Therefore, we ascribe the kink feature in tuning the carrier density to pushing this second-order $d_{xy}$ subband, denoted in the following as $d_{xy,2}$, above the Fermi level. Similar to crossing the bottom of the $d_{xz,yz}$ bands, this can be considered a Lifshitz transition. Note that this Lifshitz transition removes an electron pocket from the Fermi surface, upon increasing the carrier density.

In the Bardeen-Cooper-Schrieffer (BCS) theory, the implications of a Lifshitz transition on superconductivity in a q-2DES would be mediated through the density of states. In a q-2DES, the density of states of every subband depends stepwise on the energy. Therefore, crossing a Lifshitz transition would abruptly change the density of states at the Fermi level. If all carriers contribute equally to superconductivity, the BCS theory predicts an equally abrupt change of $T_c$ in this case. Instead, $T_c$ evolves smoothly with gate voltage, also across the $d_{xz,yz}$ and the $d_{xy,2}$ Lifshitz transitions. This does not correspond to the BCS description of a superconductor with a stepwise density of states. In real systems however, the density of states may not depend perfectly stepwise on energy. For instance, in the presence of strong spin-orbit coupling (SOC), the density of states of the $d_{xz,yz}$ band minimum is smeared out, and therefore changes more smoothly with energy \cite{joshua_universal_2012}. Despite this smearing, the density of states still increases by about an order of magnitude across the $d_{xz,yz}$ Lifshitz transition in a relatively small energy range. Therefore, the smooth gate dependence of $T_c$ across the Lifshitz transitions suggests that, in a BCS scenario, not all carriers contribute equally to superconductivity.

\begin{figure}
\includegraphics[width=0.5\columnwidth]{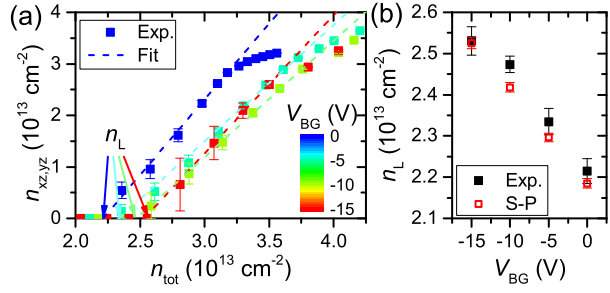}
\caption{Effect of a backgate voltage on the topgate-induced $d_{xz,yz}$ Lifshitz transition. (a) Experimentally extracted $d_{xz,yz}$ carrier density as function of total carrier density $n_{tot}$, for topgate sweeps at varying backgate voltage. The $d_{xz,yz}$ Lifshitz density $n_L$ is extracted as the total carrier density at which the linear fits to low $n_{xz,yz}$ cross the $x$-axis. (b) Extracted $d_{xz,yz}$ Lifshitz density $n_L$ versus backgate voltage. Closed symbols represent the experimental data, open symbols are results of the self-consistent Schr\"odinger-Poisson calculations.
}
\label{nltune}
\end{figure}

We now turn to the effect of a backgate voltage. Empirical modeling suggests that its primary action is to control the width of the potential well, which becomes narrower with decreasing backgate voltage \cite{bell_dominant_2009, chen_dual-gate_2016}. This should lead to an increased splitting between the energy levels of the states in the well \cite{stengel_first-principles_2011, scopigno_phase_2016, smink_gate-tunable_2017}. With a larger level splitting, more carriers can fill the $d_{xy}$ band until the Fermi level touches the $d_{xz,yz}$ band minimum. Figure \ref{nltune} shows the effect of a negative backgate voltage on $n_L$. In Fig. \ref{nltune}(a), linear fitting of the data up to the $d_{xy,2}$ transition defines $n_L$: it is the total carrier density where $n_{xz,yz}$ becomes zero. The resulting values for $n_L$ are depicted as the closed, black symbols in panel (b) and show a clear increase of $n_L$ with negative backgate voltage. This is supported by the Schr\"odinger-Poisson calculations, for which the same procedure was performed for all backgate voltages. The $n_L$ values extracted from the calculations are depicted by the open, red symbols in Fig. \ref{nltune}(b).

Besides increasing the level splitting between $d_{xy}$ and $d_{xz,yz}$ as depicted in Fig. \ref{nltune}, a narrower potential well also increases the splitting between the $d_{xy}$ subbands. This can be extracted from Fig. \ref{comp}, where $n_{xy}$ at the $d_{xy,2}$ Lifshitz transition increases with decreasing backgate voltage. We also observe that the two Lifshitz transitions are spaced closer together in topgate voltage with decreasing backgate voltage. The second-order subband is thus depleted more rapidly with stronger confinement. In the electron-electron interaction model considered here, this does not imply a change in $U$, which was taken constant across the Schr\"odinger-Poisson calculations. This suggests that in this model, the effect of the same $U$ is enhanced by a more narrow well.

The results presented above reveal that top- and backgating have profoundly different effects on the ground state of the q-2DES at the LaAlO$_3$-SrTiO$_3$ interface. Besides the previously reported disparate effect on the carrier mobility \cite{hosoda_transistor_2013}, the gating geometry also affects the band structure and superconductivity differently. In line with predictions based on band structure modeling \cite{stengel_first-principles_2011, van_heeringen_kp_2013, scopigno_phase_2016} and on previous experimental findings \cite{smink_gate-tunable_2017}, we attribute this to the effect of the changing confining potential well shape with gate voltage.

We observe that the optimal conditions for superconductivity are not necessarily coupled to a single carrier density, sheet conductivity or gate voltage. This means that SrTiO$_3$ surface states cannot be described in a universal phase diagram based on such parameters. Rather, the relative band occupation and the number of subbands contributing to transport appear to determine the electronic phase of the q-2DES. In the approximation of uncoupled, orthogonal orbitals we consider here, there are multiple subbands originating from the $d_{xy}$ orbital. The $d_{xz}$ and $d_{yz}$ orbitals have a much smaller effective mass in the out-of-plane direction and therefore, their higher-order subbands are much higher up in energy: so much higher, that the theoretical limit of 0.5 el/u.c. \cite{ohtomo_high-mobility_2004} will be reached before these subbands are populated. Therefore, in the orthogonal orbital approximation, all but two subbands contributing to transport in SrTiO$_3$ surface states are of $d_{xy}$ character.

For a full theoretical description of the system, the effects of Rashba SOC should also be taken into account \cite{caviglia_tunable_2010, joshua_universal_2012, zhong_theory_2013, diez_giant_2015, van_heeringen_kp_2013}. Rashba SOC induces interorbital coupling, giving rise to band hybridization and avoided crossings in the band structure. This results in a complex Fermi surface with large spin splittings around these avoided crossings. The occurrence of multiple subbands complicates the description of these avoided crossings \cite{van_heeringen_kp_2013,van_heeringen_theoretical_2017}. This is especially the case at weakly confined, i.e. high-mobility, interfaces, the spacing in energy between the individual subbands is small \cite{mccollam_quantum_2014} and the effect of orbital hybridization is therefore relatively strong. Despite these findings, we find a good correspondence of calculations based on uncoupled bands with the experimental results. For interfaces with a narrower potential well, we therefore propose that orbital hybridization only has a minor effect on the evolution of the band filling with gate voltage.

In summary, we have used simultaneous top- and backgating to study the relation between superconductivity and the band structure at the (001) LaAlO$_3$-SrTiO$_3$ interface. First, we revealed that the individual gate voltages affect the critical temperature differently. To understand this behavior, we mapped the evolution of the critical temperature with a combination of the two gate voltages and compared this to the corresponding gate dependence of the band filling. Besides the emergence of a second carrier type, previously established as a Lifshitz transition of the $d_{xz,yz}$ bands, we observe a second distinct feature in tuning the carrier density at higher topgate voltages. By self-consistent Schr\"odinger-Poisson calculations, we related this feature to electron-electron interactions pushing the second $d_{xy}$ subband above the Fermi level. We therefore attributed this point to a second Lifshitz transition in the subband structure of the LaAlO$_3$-SrTiO$_3$ interface. Application of a backgate voltage changes the carrier density corresponding to both Lifshitz transitions, showing that tuning the confining potential well has profound effects on the energy levels in the well.

Surprisingly, the Lifshitz transition of the second $d_{xy}$ subband correlates consistently with maximum $T_c$, thus indicating the optimal conditions for superconductivity. We therefore conclude that confinement-induced subbands are a crucial element in the phase diagram of SrTiO$_3$ surface states. Our results show that the energy levels and occupations of these subbands can be controlled electrostatically, opening numerous possibilities to harness the exotic properties of electronic subbands at the surface of complex oxides for future electronic devices.

\begin{acknowledgments}
The authors acknowledge financial support through the DESCO program of the Netherlands Organization for Scientific Research (NWO), and the European Research Council (ERC) through a Consolidator Grant.
\end{acknowledgments}

\bibliography{SC_paper}

\newpage


\begin{figure}
\includegraphics[width=0.5\columnwidth]{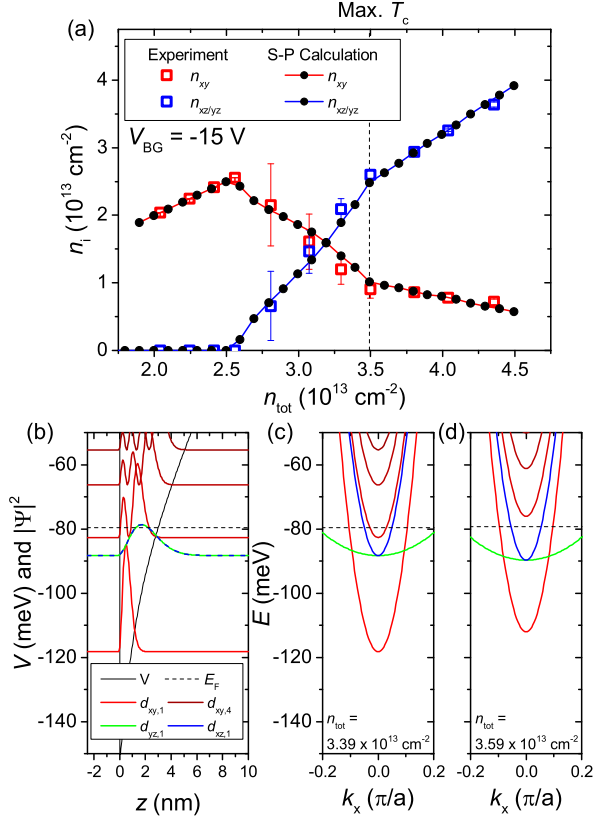}
\caption{Results of the self-consistent Schr\"odinger-Poisson calculations as function of total carrier density $n_{tot}$, for a backgate voltage of -15 V. Input parameters are discussed in the main text. (a) Comparison of measured and calculated band filling versus total carrier density. Open symbols represent the measured values, closed symbols (connected by a line as guide to the eye) depict the calculated values. The vertical dashed line indicates the experimentally found filling where $T_c$ is maximized for $V_{BG}$ = -15 V. (b) Self-consistently calculated potential well for a total carrier density of 3.39 $\times$ $10^{13}$ cm$^{-2}$, with probability functions $|\Psi|^2$ indicated within the well for each subband. The displayed potential $V$ corresponds to the $d_{xy,1}$ subband, the effective potentials for the other bands differ from this one by a few meV through the effective interaction model. Energies are defined relative to the SrTiO$_3$ bulk conduction band. (c) Calculated subband dispersion corresponding to the potential well in (b), at $n_{tot}$ = 3.39 $\times$ $10^{13}$ cm$^{-2}$: just below the filling corresponding to maximum $T_c$. (d) Same as (c), for a total carrier density of 3.59 $\times$ $10^{13}$ cm$^{-2}$: just above the filling corresponding to maximum $T_c$. The legend in (b) applies to (c) and (d) as well.
}
\label{SP_comp}
\end{figure}

\end{document}